# Order Handling in Convergent Environments

Jordan Vrtanoski
BSS CU GCC, RMEA
Ericsson AB
Downtown Jebel Ali, Dubai, AUE
jordan.vrtanoski@ericsson.com

Toni Stojanovski
Faculty of informatics
European University of R. Macedonia
Skopje, Macedonia
toni.stojanovski@eurm.edu.mk

*Abstract –* **The rapid development of IT&T technology had big impact on the traditional telecommunications market, transforming it from monopolistic market to highly competitive high-tech market where new services are required to be created frequently. This paper aims to describe a design approach that puts order management process (as part of enterprise application integration) in function of rapid service creation. In the text we will present a framework for collaborative order handling supporting convergent services. The design splits the order handling processes in convergent environments in three business process groups: order capture, order management and order fulfillment. The paper establishes abstract framework for order handling and provides design guidelines for transaction handling implementation based on the checkpoint and inverse command strategy. The proposed design approach is based in a convergent telecommunication environment. Same principles are applicable in solving problems of collaboration in function of order processing in any given heterogeneous environment.**

## I. Introduction

Rapid development in the field of IT&T had transformed the traditional monopolistic telecommunications market to highly competitive environment. At the end of the 20[th] century, in the 80's and beginning of the 90's, installation of a new telecommunications service was measured in months or weeks. The process of new customer acquisition was at the bottom of the priority list. The main income of the telecom operator at the end of 20[th] century was the voice traffic. The digital communication was present mainly in the domain of big enterprises.

The beginning of the 21[st] century brought the expansion of the Internet. The digital communication was no longer in the domain of the enterprises only, but became available and affordable for the population. The number of Internet users increased from 2,6 millions in 1990, to 1,5 billion in 2008[1].

Increased amount of Internet users allowed for development of alternative telecommunication services like voice over IP, video chat, web hosting, etc. For the first time there was viable alternative to the telephony (analog voice) as a service. Soon enough it was obvious that the data services / data traffic is the future of the telecommunications industry [2].

Adding to this, the deregulation of the telecom operators that began in mid 90's brought the competition in the telecommunication market. The awareness of the operator for the necessity of the attitude change [3] was evermore present.

First decade of the 21[st] century brought new shift to the paradigm of the post *dotcom* operators. Average customer had 4-5 services, each one of them sold, billed and supported separately. This caused increased operational cost on the operator side, as well as decreased customer satisfaction by the fact that the customer had different interaction (depending on the service) with the same operator. The paradigm of convergent services was a necessity in order to address the issues [4]-[6]

As all other IT&T systems at the operators suffered changes in order to become compliant with the service convergence, also the selling, provisioning and billing support systems were affected.

The convergent telecom environment represents a heterogeneous environment that consists of IT and telecommunication hardware and applications. The applications in majority of the cases are vendor specific and are exposing vendor specific application programming interfaces.

The convergence of the services requires customer experience to be consistent (as much as the technical nature of the service allows) across all services that the customer can obtain from the operator. In this work we describe a design approach that puts order management process in function of rapid service creation, and present a framework for collaborative order handling supporting convergent services.

Here is the overview of the paper. In Section I of this paper we provide overview of the evolution of convergent environment. Section II aims to define the problem of order handling in the convergent environments and to lay down basic requirements for the design. Section III describes the proposed design approach. Section IV gives high-level overview of interaction between the components of the proposed design model. Section V gives the concluding remarks and directions for future work.

## II. ORDER HANDLING PROBLEM

Order handling represents a group of processes that sits between the customer who generates the demand on the top, and the service providers at the bottom, which are serving the demand. The service provider, in this case, represents a service platform that belongs to the operator or to a third party.

At the top, the customer can interact with the operator (trough it's distribution channels) to express his will to purchase a product which will subscribe him to a given service provided by the service provider (the operator or business partner). The interaction between the customer and the operator can be performed via multiple distribution channels such as point of sale; IVR; WEB or CSR. Regardless of the distribution channel chosen by the customer to purchase the product, the customer's experience should remain unchanged.

Columbus in his study [7] identified that often companies have different order capture and management system for different distribution channels, and that this situation directly translates in confusion of the demand. Columbus concludes that successful companies have implemented multichannel order handling system (and processes).

Products that are offered to the customers can contain one or more services. Lately, telecom operators are creating multi-play offers by bundling different products, by example broadband subscription with telephony subscription. Order management must be flexible enough in order to support decomposition of product to multiple services.

Support for the convergent services brings multiple service platforms in the picture. Though the services offered by the convergent operator have the same look and fill, the services are supported by different service platforms, often belonging to entities outside the operator itself. This adds complexity to the design of the order management systems in order to support the convergent service delivery in the standard Operations Support Systems/Business Support Systems (OSS/BSS) landscape[8].

Successful order handling would require implementation of automation mechanisms that:
- Will support multiple distribution channels with different technical/commercial demands;
- Will have high reusability of the automated business tasks;
- Will support multiple service platforms;
- Will support human to human (H2H), human to system (H2S) and system to system (S2S) interaction;
- Will support decomposition of product to services;
- Will support orders containing products/services that will be fulfilled at different service platforms (convergent services);
- Will allow seamless introduction of new distribution channels and/or service platforms and
- Will allow seamless replacement of the service platforms and seamless cease of the distribution cannel.

## III. DESIGN APPROACH

In order to facilitate multi-channel support by the order handling system, the order handling processes should be split in three main groups: order capture processes; order management processes and order fulfillment processes. Consequently, the order handling system should consist of three main modules: order capture module; order management module and order fulfillment module. The process groups are shown on the Figure 1.

By decoupling the order capture from the order management, the system will have flexibility to support variety of distribution channels without necessity of adopting internal logic to cope with the technology of the distribution channel. Introducing, modifying or ceasing a distribution channel will have no impact on the management and fulfillment of the orders.

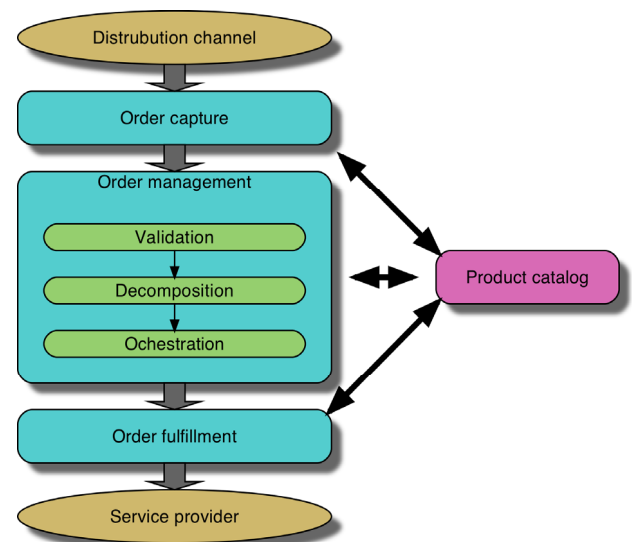

Figure. 1. Order handling processes

Additionally, by decoupling the order management from order fulfillment, the system gains further flexibility. The order handling system will be able to involve multiple service platforms regardless of the technology of the platform, location and governance. The order fulfillment module will be responsible to adopt the order management messages to the service platform's requirements.

### A. Order capture

Order capture processes are responsible for interaction with the customer through the distribution channel. The processes capture customer's request for products (services and goods) that the operator provides.

The order capture module should provide standard framework for the interface that will accept the order from the user. The interface will depend from the channels nature. By example, the IVR interface will be in a form of voice menus; the Web interface will be in a form of HTML pages; etc.

The user of the module can be the customer (cases with IVR, web, etc.) or customer representative (call center, point of sale, etc.).

The module should be built following the MVC pattern [9]. The model and the controller will be re-usable components of the order capture framework, while the view will depend on the technology that represents the distribution channel.

The approach described here is same regardless if the distribution channel is in control of an operator (POS, IVR, etc.) or in control of a business partner (sales dealer, service enabler, etc.).

*B. Order management*

Order management processes are responsible for validation of the order; splitting and distributing the order toward service platforms; proper sequencing of the service platform invocation and collaboration.

The order management module should provide mechanisms for order validation; order decomposition and order orchestration.

**Order validation** submodule should provide library of business and technical rules that can be used to determine if an order is valid and should be accepted for further processing. For example, a business rule can be the customer's credit limit or the customer's contractual terms; a technical rule can be service availability in the customer's premises or customer-premises-equipment (CPE) capabilities.

**Order decomposition** submodule will provide mechanism to generate multiple sub-orders based on the product definition (product decomposition) or based on the fulfillment destination (work order decomposition). By example, a given product that is requested by the customer consists of broadband service and telephony service. This will trigger fulfillment in two service platforms and in the billing platform, consequently the decomposition will produce tree sub-orders. Each of the service platforms needs only a part of the product information that is related to the service that the platform provides, while the billing platform will require all billable services.

**Order orchestration** submodule will provide a mechanisms to coordinate the order fulfillment by resolving the inter-dependences between sub-orders and by coordination of the subsequent fulfillment processes. As in the previous example, a set of given services composing a product will be fulfilled (provisioned) on the respective service platforms, and only after successful fulfillment on all platforms, the product will be provisioned on the customer's financial account in the billing system.

Orchestrating the fulfillment across multiple platforms will require transaction management capabilities to be implemented in the order orchestration submodule. Transaction handling is discussed later in this text.

The order decomposition submodule and the order orchestration submodule will generate dynamic fulfillment plan.

*C. Order fulfillment*

Order fulfillment processes are responsible for interaction with the service platforms and users in function of providing the service/goods to the end user.

The order fulfillment module represents abstraction layer for the legacy systems (service platforms and business support systems). Each of the legacy systems represents a fulfillment target from the perspective of order handling.

The module is based on a workflow engine as suggested by Ouyang et al. [10] and Lin and Chang [11] in their work. Service platforms that will interact in the fulfillment process are represented as business services using resource adapter pattern [12].

In some cases, the service platform invoked in the order fulfillment belongs to a business partner. The business partner that provides the services to the operator becomes member of the operator's supply chain. Facilitating order fulfillment in the supply chain requires B2B communication. Most often case is by utilizing B2B gateways[13].

*D. Transaction handling*

Transaction in the case of order handling is distributed across multiple systems (service platforms). Diversity of the platforms adds to the complexity of the transaction handling. Additionally, the order fulfillment processes can be semi automated, in which case we speak about long-lived orders. Fulfillment of the order in majority of the cases will require transaction (results of the previous fulfillment steps) to be committed and changes to be visible in the service platforms. For example, scheduling the visit to the customer premises in the workforce management system requires address of the customer's premises to be committed in the CRM system.

The order management module, more specifically the order orchestration submodule, and the order fulfillment module should provide mechanisms for transaction handling across service platforms and business support systems.

We suggest the transaction handling mechanism to be based on non-linear undo buffer. The non-linear undo buffer can be achieved either by implementing command pattern to achieve inverse command strategy [14] or by implementing memento pattern to achieve checkpoint strategy [14].

Due to the complexity of the order handling, our suggestion is order handling to provide implementation for both patterns and flexible mechanism for selection which undo approach to be taken per given business scenario (given use case).

*E. Product catalog*

All three main components of the order handling process are dependant of the electronic product catalog [15]in order to facilitate better communication.

As discussed, the convergent order handling enables the customer to purchase products and services from multiple service platforms trough a given distribution channel.

The information about the products offered trough a distribution channel should be unified. Lohse and Schmid are suggesting implementation of Mediated Electronic Product Catalog (MEPC)[16].

MEPC should provide high-level view (customer's view) of the products. During the order management and fulfillment, the product will be decomposed on multiple products and services offered by the individual service platforms.

*F. Human Colaboration*

Handling the orders in the convergent environment requires human interaction in form of human-to-human and human-to-machine interaction.

The framework proposed in this paper can be utilized to facilitate these two types of interaction in two ways:

- interaction can be modeled as serious of human tasks performed during the orchestration of the order. In this case, each instance of the interaction will be represented as a sub-order;
- alternative approach to modeling human interaction is by integrating a dedicated task management system. In this case the human interaction will be performed in the boundaries of the dedicated system.

In addition, each service platform can contain it's own implementation of the human-to-computer interaction. The human-to-computer interaction inside the fulfillment target boundaries is permitted.

## IV. INTERACTION BETWEEN COMPONENTS

We suggest that the interaction between the modules (shown on Figure 2) is based on standardized protocols. We also suggest this protocol to be based on JMS messages with XML payload. JMS messaging provides the ability to implement asynchronous and synchronous calls within the same interface. In the further text we will present the collaboration patterns based on the JMS interfaces.

The order capture module should prepare the XML message of the order and should submit the message to the JMS queue of order manager component.

The order manager will pick up the message and will validate its content based on the pre-configured validation rules. If the XML message represents a valid order, then the order will be decomposed into sub-orders.

Orchestration sub-module will traverse through the collection of the sub-orders and will resolve the dependencies. This will produce the execution plan. Once the cross dependencies are resolved and the execution plan is created, the orchestration sub-module picks up a sub-order from the sorted set and submits it to the fulfillment module.

Fulfillment module translates the sub-order into list of actions that will be executed on the legacy system. Possible case is that some of the actions to be performed require user intervention. Such is a case in which the CPE should be installed or replaced. This will require field technician to visit customer premises before the fulfillment can continue. For simplicity the actors are also abstracted as a legacy system executing manual actions.

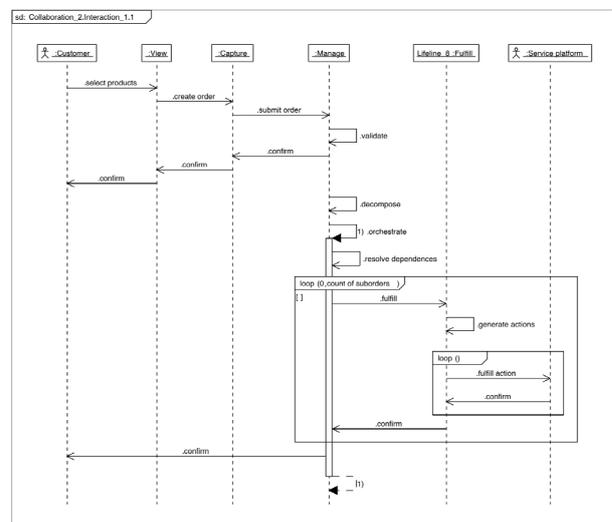

Figure. 2. Interaction between components.

Once the list of actions is successfully applied on the legacy system, the orchestration stores the result and appends the remaining sub-orders with the data returned by the previous fulfillment, by example, the MAC address of the CPE installed in the customer premises is required by the voice platform.

## V. CONCLUSION

In this paper, we have presented design approach in facilitating automation of order handling process. The design allows for greater level of flexibility in order handling across multiple distribution channels and multiple service platforms. Design allows distribution channels and service platforms to be placed outside the operator's infrastructure.

The problem of order handling in convergent environments was successfully addressed by decoupling of order capture from order management; and order management from order fulfillment. This split allowed introduction of standardized interfaces between the three components.

Although the presented design approach is suggested for implementation in telecommunication enterprise, the same design approach can be implemented in any other enterprise having heterogeneous distribution network and/or heterogeneous fulfillment process.

The NGOSS (New Generation of Operations Systems and Software) models proposed by "TM Forum" are becoming more popular between enterprise architects in telecommunication domain. Lately, the standards proposed in the scope of "TM Forum Frameworx" (formerly known as NGOSS) [17] became paramount requirements for IT components in the telecommunication environments.

Future work will be done to propose data models for each of the proposed modules (Capture, Management and Fulfillment). The data models will be aligned with SID [18] data model proposed by "TM Forum" group. Upon definition of the data model, interfaces between the components should be standardized.

Order handling processes depend on the information stored in the inventory platforms, such as service inventory, network inventory, etc. Future work should be done do define this interaction as a set of standardized interfaces.